\documentstyle[epsfig,psfig]{mn}

\newcommand{\gtae}{$\buildrel {\lower3pt\hbox{$>$}} \over 
{\lower2pt\hbox{$\sim$}} $}
\newcommand{\ltae}{$\buildrel {\lower3pt\hbox{$<$}} \over
{\lower2pt\hbox{$\sim$}} $}

\title[An irradiated accretion disk in RE~J1034$+$396]
    {An irradiated accretion disk in the narrow-line Seyfert 1 RE~J1034$+$396?}

\author[R. Soria \& E. M. Puchnarewicz]
{R. Soria$^{1}$ and E. M. Puchnarewicz$^{1}$ \\ 
$^{1}$Mullard Space Science Laboratory, 
  University College London, Holmbury St. Mary, Dorking,
  RH5~6NT; email: rs1@mssl.ucl.ac.uk}

\date{Received: }

\begin{document} 

\def\1034{RE~J1034$+$396} 
\def\Mdot{\hbox{$\dot M$}}
\def\Msun{\hbox{M$_\odot$}}
\def\Rsun{\hbox{$R_\odot$}} 
\outer\def\gtae {$\buildrel {\lower3pt\hbox{$>$}} \over 
{\lower2pt\hbox{$\sim$}} $}
\outer\def\ltae {$\buildrel {\lower3pt\hbox{$<$}} \over
{\lower2pt\hbox{$\sim$}} $}
\def\rchi{{${\chi}_{\nu}^{2}$}} 
\def\simlt{\lower.5ex\hbox{$\; \buildrel < \over \sim \;$}}
\def\simgt{\lower.5ex\hbox{$\; \buildrel > \over \sim \;$}}

\maketitle

\begin{abstract}
We model the optical to X-ray continuum spectrum of the narrow-line 
Seyfert 1 galaxy \1034. We show that the flat optical spectrum 
is consistent with emission from an irradiated accretion disk. 
The X-ray emission can be modelled with a disk blackbody 
and a Comptonised component. The temperature at the inner edge 
of the disk $T_{\rm in} = (0.12 \pm 0.02)$ keV. Using this constraint, 
we show that the flat optical spectrum 
is consistent with emission from the irradiatively heated 
outer part of the accretion disk. We constrain the outer radius 
of the optically thick disk ($R \simgt 5 \times 10^{16}$ cm) 
and the inner radius of the irradiation-dominated region 
($R \simgt 5 \times 10^{12}$ cm). 
Our optical and X-ray spectral fits indicate a mass 
$0.6 \simlt M \simlt 3 \times 10^6 \Msun$, 
and do not rule out a low (i.e. face-on) inclination angle 
for the system.

\end{abstract}

\begin{keywords}
    accretion, accretion discs --- galaxies: active --- 
     galaxies: nuclei --- galaxies: Seyfert  --- 
    galaxies: individual: \1034 --- X-rays: galaxies  
\end{keywords}

\section{Introduction}

The optical to X-ray continuum of \1034 is highly unusual 
for a Seyfert 1 galaxy. The optical/UV continua of most AGN rise 
towards the blue with a slope $\alpha \approx 0.4$ 
(where $\alpha$ is the spectral index, defined 
such that $F_{\nu} \propto \nu^{-\alpha}$), while 
the soft X-ray spectrum falls towards high energies 
with spectral index $\alpha \approx 2$ (e.g.\ Laor et al.\ 1997). 
The spectrum seems to peak in the unobservable EUV 
and this continuous, optical to soft X-ray feature, 
known as the ``big blue bump'', is believed to represent 
the emission from a geometrically thin, optically thick accretion disk.

In \1034, however, the optical/UV continuum is flat ($\alpha \approx 1$), 
with no sign of the big blue bump down to Ly$\alpha$. 
At $\sim 0.1$ keV, the soft X-ray spectrum is very strong above 
the extrapolated level of the optical/UV continuum, 
peaking at $\approx 0.3$ keV then falls steeply towards higher
energies (Puchnarewicz et al.\ 1998). 
Puchnarewicz et al.\ (2001) suggested that the accretion disk component 
in \1034 is one of the hottest observed in AGN, so that 
its high-energy turnover is measurable in the soft X-rays 
rather than the UV band. They inferred a 
high accretion rate ($L \approx 0.3$--$0.7$ $L_{\rm Edd}$), 
a small black hole (BH) mass $M \approx 10^6 \Msun$ 
and an inclination angle $i \approx 60$--$70^{\circ}$.

The flat optical spectrum was modelled by Puchnarewicz et al.\ (2001) 
with an underlying power-law component extending
from the optical to hard X-rays, in addition 
to the optically thick disk spectrum. However, they could find 
no satisfactory physical explanation for this feature. 
Breeveld \& Puchnarewicz (1998) detected no significant polarization 
in the optical, making a synchrotron origin unlikely. 

In this Letter, we suggest a different 
interpretation of the continuum spectrum: we show 
that the flat optical spectrum is consistent 
with the continuum emission of an irradiated disk, with no need to invoke  
an additional power-law component. We also revise the estimate 
for the central mass, and note that low inclination 
viewing angles cannot be ruled out.

\section{Observations}

%\subsection{Optical: William Herschel Telescope}

Optical spectra of \1034 were obtained on 1996 March 24 with the ISIS
spectrograph mounted on the William Herschel Telescope at La Palma (in
service observing mode). The spectrum, which covered the range
3800--7000 \AA, is clearly contaminated by light from the host galaxy,
particularly in the red.  In order to remove the extended galactic
emission, the nuclear spectrum was deconvolved from the total spectrum
using the technique described in Puchnarewicz et al.\ (2001).

%\subsection{UV: Hubble Space Telescope}

In the UV band, \1034 was observed by {\it HST} on 1997 January 31 
using three gratings (G130L, G190L and G270L) covering 
the range 1100--3300 \AA. Full details of the observations 
and data reduction were presented in Puchnarewicz et al.\ (1998).

%\subsection{X-rays: {\it Beppo-SAX}}

{\it Beppo-SAX} observed the system on 1997 April 18. 
Useful data were obtained with the Low Energy Concentrator Spectrometer 
(LECS; 0.1--10 keV) and the Medium Energy Concentrator Spectrometer 
(MECS; 1.3--10 keV); all three MECS units were used. 
The source was not detected 
with the high-energy instruments.
The total observing time was 21.7 ks for LECS and 43.2 ks for MECS. 
For more details on the observations and on the techniques 
used for source extraction and background subtraction, see 
Puchnarewicz et al.\ (2001).

\begin{figure}
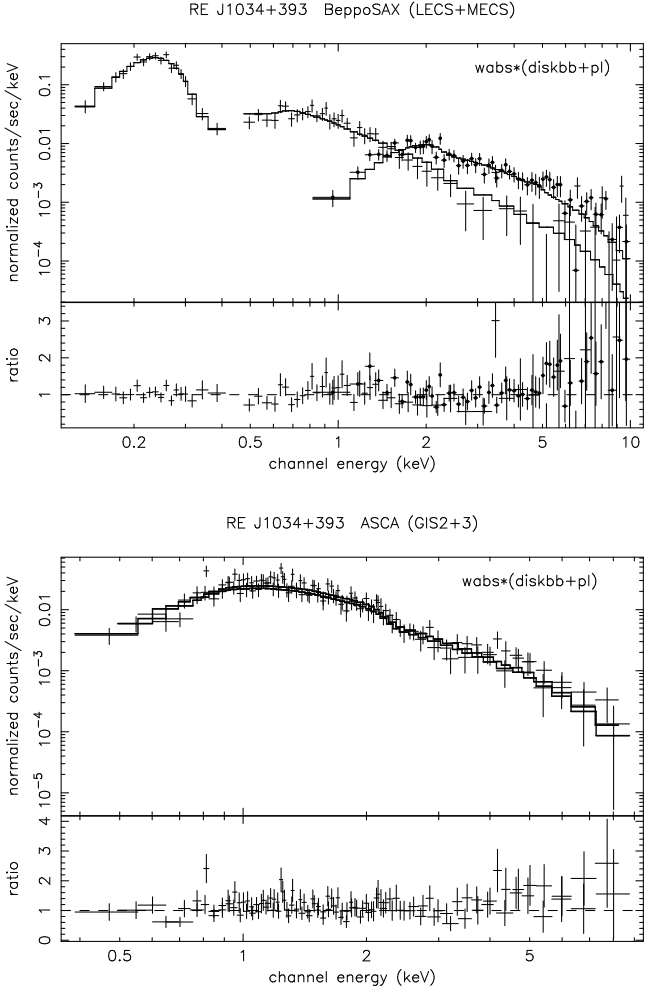

\begin{center}   
\epsfig{figure=fit1b.ps, width=6.3cm, angle=270}\\

\vspace{0.5cm}   

\epsfig{figure=fitgis1b.ps, width=6.3cm, angle=270} 
\end{center}    
\caption{Top panel: disk blackbody plus power law fit 
to the 0.15--10 keV {\it Beppo-SAX} spectrum. The fit parameters are given 
in Table 1. Bottom panel: the same disk blackbody plus power law model 
(parameters fixed as in Table 1) 
applied to the 0.4--8 keV {\it ASCA} spectrum.}
\end{figure}

\begin{table}
%\begin{table*}
%\begin{minipage}{150mm}
\begin{minipage}{80mm}
\caption{{\small XSPEC} fitting parameters to the {\it Beppo-SAX} data}
\label{mathmode} 
\centering 
\begin{tabular}{@{}lr} 
\hline
\hline
Parameter &  Value \\ 
\hline
\multicolumn{2}{c}{wabs*(diskbb + power law): $\chi_\nu^2 = 1.00$ (121 dof)}\\
\hline
$n_{\rm H}$ (column density)     & $1.5 \times 10^{20}$ cm$^{-2}$ \\[5pt]
$T_{\rm in}$ ($T$ at inner disk edge)   
	& $0.116^{+0.007}_{-0.008}$ keV \\[5pt]             
$K_{\rm dbb}$ (diskbb normalisation)  & $3873^{+1368}_{-1068}$ \\[5pt] 
$\Gamma$ (photon index)       & $2.59^{+0.16}_{-0.14}$ \\[5pt]
$K_{\rm pl}$ (power-law normalisation)   & $7.03^{+1.14}_{-1.00} \times 10^{-4}$ \\ [5pt]   
\hline
\multicolumn{2}{c}{wabs*(diskbb + comptt): $\chi_\nu^2 = 1.05$  (119 dof)}\\
\hline
$n_{\rm H}$  (column density)   & $1.5 \times 10^{20}$ cm$^{-2}$ \\[5pt]
$T_{\rm in}$ ($T$ at inner disk edge)   & $0.116^{+0.007}_{-0.009}$ keV \\[5pt]
$K_{\rm dbb}$ (diskbb normalisation)  & $4101^{+1323}_{-964}$ \\[5pt]
$T_{0}$ (input soft photon $T$)        & $0.05^{+0.05}_{-0.05}$ keV \\[5pt]
$T_{\rm c}$ ($T$ of hot corona)     & $14.8$ keV \\[5pt]
$\tau$ (optical depth of hot corona)         & $1.307$ \\[5pt]
$N_{\rm c}$ (normalisation)    & $3.54 \times 10^{-3}$ \\
\hline
\end{tabular}   
\end{minipage} 
%\end{table*}  
\end{table}

\section{X-ray spectral analysis}

In Puchnarewicz et al.\ (2001), LECS and MECS spectra 
were fitted with a number of models (see their Table 4) 
including power-law and single-temperature blackbody components. 
The best fit to the soft spectrum was obtained 
with two blackbody components and a power law, with a column density 
fixed at the Galactic value ($n_H = 1.5 \times 10^{20}$ cm$^{-2}$). 

In this work, we want to test our hypothesis that 
the optical-to-X-ray continuum spectrum is dominated 
by an optically thick accretion disk. This would be analogous to 
Galactic BH candidates in the soft spectral state.
Therefore, we fitted the {\it Beppo-SAX} spectra 
with a multi-temperature disk blackbody plus a power law, 
using {\small XSPEC} version 11.0.1 (Arnaud 1996).
This phenomenological model can be taken as an approximation of the spectrum 
produced by an accretion disk plus a hot Comptonising region. 
The disk is responsible for the soft emission; part of the soft photons 
are inverse-Compton scattered to the hard X-ray band as they cross 
the hot corona. We also fitted the data with the more physical 
Comptonisation model of Titarchuk (1994) (comptt in {\small XSPEC}).

The best-fit parameters for the two models 
are listed in Table 1, and the spectrum fitted with 
disk blackbody plus power law is shown 
in Figure 1 (top). 
We fitted the LECS and MECS data simultaneously 
(using 0.13--10 keV for LECS, and 0.9--10 keV for MECS), 
and we applied a constant factor of 0.9 in the relative 
normalisation of the two instruments (Fiore et al.\ 1998).
For both the models presented here, the best-fit 
intrinsic column density converged to 0; we have therefore 
fixed the column density to the Galactic value.
For the disk plus power-law model, the
power-law spectral index $\alpha = 1.6 \pm 0.2$ 
(photon index $\Gamma = 2.6$) 
is softer than for typical AGN (as previously 
suggested by Pounds, Done \& Osborne 1995).

Puchnarewicz et al.\ (2001) claimed that  
there is firm evidence of a hardening in the spectrum 
at energies $\simgt 7$ keV; they suggested that a harder 
power-law component dominates at higher energies, 
and that the same power-law component may explain the optical spectrum 
if it extends to those lower energies.
A hint of an upturn in the spectrum for $E \simgt 7$ keV 
is indeed visible in Figure 1. However, given the low signal-to-noise 
ratio in the higher energy channels, 
and the fact that the upturn is evident only in the LECS data 
(less reliable at those energies), we prefer not to speculate 
on the physical reasons for the possible existence of a hard 
component. 

We also used the best-fit parameters inferred from the 
{\it Beppo-SAX} data (Table 1) to re-fit the {\it ASCA}/GIS data 
from 1994 November, 
previously discussed in Pounds et al.\ (1997) and Puchnarewicz 
et al.\ (2001). The result is shown in Figure 1 (bottom); we obtain 
$\chi^2_\nu = 0.86$ (117 d.o.f.) for the disk blackbody plus 
power-law model. 
Again, although there may be marginal evidence for a hardening 
of the spectrum at $E \simgt 5$ keV, the signal-to-noise 
ratio is too low to allow a meaningful interpretation.

A significant result from both models is the strong constraint 
on the temperature at the inner radius of the accretion disk. 
We obtain $kT_{\rm in} = 0.116^{+0.07}_{-0.09}$ keV. 
This is also in agreement with earlier {\it ROSAT} 
(Puchnarewicz et al.\ 1998) and {\it ASCA} (Pounds et al.\ 1997; 
Puchnarewicz et al.\ 2001) observations. We shall use this firm 
result as we model the disk spectrum 
in the optical band.

\section{Continuum spectrum of an irradiated disk}

The effective temperature of an accretion disk photosphere at each radius 
is determined by the thermal energy generated by local viscous dissipation 
near the disk mid-plane (viscous heating), 
as well as by the energy intercepted 
from the central X-ray source, thermalised and re-emitted 
(irradiative heating). 

In the absence of irradiation,  
the effective temperature of a geometrically thin,  
optically thick disk is (Shakura \& Sunyaev 1973; see also, e.g., 
Frank, King \& Raine 1992)

\begin{equation}
T_{\rm v}^4(R) = \left (\frac{3GM\dot{M}}{8 \sigma \pi R^3}\right )^{1/4} \, 
	\left [1- \left( \frac{R_{\rm c}}{R} \right )^{1/2} \right ]^{1/4},
\end{equation}
where $R_{\rm c}$ is a critical radius ($\leq R_{\rm in}$) 
beyond which we assume no 
further viscous dissipation in the accretion flow, 
and $\dot{M}$ is the mass accretion rate 
at radius $R$.

The contribution of the irradiation to the effective temperature 
is (e.g., King 1998):

\begin{equation}
T_{\rm irr}^4(R) 
= \frac{ \eta \dot{M}_{\rm c} c^2 (1-\beta)}{4 \pi \sigma R^2} 
	\, \left (\frac{H}{R} \right )^n  \left ( \frac{{\rm d} \ln H}
	{{\rm d} \ln R} -1 \right ),
\end{equation}
where $\eta$ is the efficiency of conversion of accreted mass 
into energy, $\dot{M}_{\rm c}$ is the central mass accretion rate 
($\dot{M}_{\rm c} < \dot{M}(R)$ if some mass is lost in a disk wind), 
$\beta$ is the albedo of the disk surface, $H$ is the disk scale-height 
at radius $R$. The index $n$ ($1 \leq n \leq 2$) 
depends on the geometry of accretion 
(Vrtilek et al.\ 1990; van Paradijs 1996).

The last two factors on the right hand side of equation (2) depend 
on the disk thickness and the amount of disk flaring 
(a more concave disk intercepts a larger fraction of radiation). 
They vary only slowly with $R$; following King (1998), we approximate them 
with a constant factor $\gamma \equiv (H/R)^n ({\rm d} \ln H /{\rm d} 
\ln R -1)$, with $0.01 \simlt \gamma \simlt 0.05$ for a thin disk. 
Hence, $T_{\rm irr} \sim R^{-1/2}$ while $T_{\rm v} \sim R^{-3/4}$; this 
implies that irradiative heating dominates at large radii, and viscous 
heating at small radii.

Henceforth, we shall also assume that $\dot{M}_{\rm c} = \dot{M}(R)$, 
and re-write the mass accretion rate in terms of the emitted 
luminosity: $\dot{M}(R) = (L/L_{\rm Edd}) \, (L_{\rm Edd}/\eta c^2)$ 
$\equiv l_{\rm Edd} \, (L_{\rm Edd}/\eta c^2)$, with 
$L_{\rm Edd} = 1.3 \times 10^{38} \, (M/M_{\odot})$ erg s$^{-1}$. 
Finally, given the characteristic dimensions of \1034, 
we shall define $r_{12} \equiv R/(10^{12} {\rm cm})$ 
and $m_{6} \equiv M/(10^6 M_{\odot})$.

From equations (1) and (2), 
the total effective temperature is given by:

\begin{equation}
T_{\rm eff}(R) =  \left [ T^4_{\rm v}(R) + T^4_{\rm irr}(R) \right ]^{1/4} 
 \simeq  a r_{12}^{-3/4} \left [ 1 + \frac{r_{12}}{l} \right ]^{1/4},
\end{equation}
with the parameters $a$ and $l$ given by:

\begin{equation}
a \simeq 6.7 \times 10^5 \left (\frac{0.2}{\eta}\right)^{1/4} 
l_{\rm Edd}^{1/4} \, m_{6}^{1/2} \ {\rm K}
\end{equation}
\begin{equation}
l^{-1} \simeq 1.4 \times 10^{-2} \left (\frac{\eta}{0.2}\right) 
\left ( \frac{1-\beta}{0.5} \right)
\left ( \frac{\gamma}{0.03} \right) m_6^{-1}.
\end{equation}
Hence, we have

\begin{equation}
T_{\rm in} \equiv T_{\rm eff}(R_{\rm in}) 
\simeq a \left (\frac{R_{\rm in}}{10^{12}\, {\rm cm}}\right )^{-3/4} 
\equiv a r_{\rm in}^{-3/4}.
\end{equation}

The specific flux at frequency $\nu$ detected  
by an observer at distance $d$, with the line of sight 
forming an angle $i$ to the normal to the disk plane, 
is (Mitsuda et al.\ 1984): 

\begin{equation}
F_{\nu} = \frac{4 \pi h (\cos i) \nu^3}{c^2 d^2 (1+z)^2} 
	\int_{R_{\rm in}}^{R_{\rm out}}
	\frac{R\, {\rm d}R}{\exp{(h\nu/kT_{\rm eff})} -1},
\end{equation}
where the redshift $1 + z = 1.042$ for \1034. This corresponds 
to a proper distance $d = 3.8 \times 10^{26} h^{-1}$ cm. Here 
we have defined, as usual, $h = H_0/(100$ km s$^{-1}$ Mpc$^{-1})$.

Equation (7) is sometimes modified with the introduction 
of a spectral hardening factor $f$ (Ebisuzaki, Hanawa \& Sugimoto 1984; 
Shimura \& Takahara 1995), to take into account Compton scattering 
in the disk. We can write the ``diluted disk blackbody'' specific flux as:

\begin{equation}
F_{\nu}^{\rm db} = f^{-4} F_{\nu}(fT_{\rm eff})
\end{equation}
with $1 \leq f \simlt 1.7$ (Shimura \& Takahara 1995)

Our goal is to check if a disk blackbody or a diluted 
disk blackbody spectrum can explain the observed optical/UV/soft 
X-ray continuum. We have compared the observed 
fluxes $\nu F_{\nu}$
with the values predicted by equations (7) and (8), 
for various values of the parameters $r_{\rm in}$, $r_{\rm out}$, 
$a$, $l$ and $i$. 
Using the results of our X-ray spectral fitting, we constrain   
$T_{\rm in}$ to vary in the range $0.10 \leq kT_{\rm in} \leq 0.15$ keV. 
We shall discuss our main results in the next section.

\begin{table}
%\begin{table*}
%\begin{minipage}{150mm}
\begin{minipage}{80mm}
\caption{Best-fit parameters for the irradiated disk model 
[see Eqs.\ (3), (7), (8)]}
\label{mathmode} 
\centering 
\begin{tabular}{@{}lr} 
\hline
\hline
Parameter &  Value \\ 
\hline
$f$             & 1.5 \\[5pt]
$a$             & $7.5 \times 10^{5}$ K\\[5pt]
$l$             & 43.5 \\[5pt]
$R_{\rm in}$    & $0.56 \times 10^{12}$ cm \\[5pt]             
$R_{\rm out}$   & $5.0 \times 10^{16}$ cm \\[5pt] 
$h^{-2} \cos i$ & $0.235$ \\
\hline
\end{tabular}   
\end{minipage} 
%\end{table*}  
\end{table}

\section{Discussion of our results}

We show in Figure 2 that the continuum spectrum at energies $\simlt 0.3$ keV 
can be well fitted by a diluted disk blackbody spectrum (thick solid 
line) with $f = 1.5$, and best-fit parameters listed in Table 2. 
For $E \simgt 0.5$ keV, the thick solid line plotted 
in Figure 2 represents the best fit to the {\it Beppo-SAX} 
X-ray data, discussed in Section 3. 

The model parameters in Table 2 imply $kT_{\rm in} = 0.10$ keV.  For
$h=0.5$, it is $i = 20^{\circ}$, while for $h=0.75$, $i =
60^{\circ}$. Therefore, unlike what was claimed in Puchnarewicz et
al.\ (2001), we cannot rule out that the system is seen face-on ($i
\simlt 45^{\circ}$).  This would explain why the molecular torus does
not block out the X-ray emission from the central source. It has been
suggested (Puchnarewicz et al.\ 1992; Boller et al.\ 1997) that
narrow-line Seyfert 1 galaxies are generally seen face-on.

If we assume a simple disk blackbody spectrum instead ($f=1$), 
it is clear from 
equations (7) and (8) that the same fit can be obtained 
with $a = 1.1 \times 10^6$ K, $h^2 \cos i = 0.04$ ($i = 80^{\circ}$ 
for $h=0.5$), with the other parameters fixed as above. 
This corresponds to $kT_{\rm in} = 0.15$ keV. 
In fact, we can obtain good fits for $0.10 \simlt kT_{\rm in} 
\simlt 0.15$ keV by varying $f$ between 1.5 and 1. 
Given the uncertainty in the distance 
and in the hardening factor, we cannot constrain 
the inclination angle more strongly.

The parameter $l$ provides the size of the disk region 
dominated by irradiative heating. In our best fit, 
the disk is irradiatively dominated for $R \simgt 80 \, R_{\rm in}$. 
Small changes in the parameter $l$ have large 
effects on the predicted optical spectrum. In Figure 2, we also plot 
the spectrum obtained with $l = 60$ (marked with ``b''; irradiation 
dominating for $R \simgt 110 \, R_{\rm in}$) and $l = 30$
(marked with ``c''; irradiation 
dominating for $R \simgt 50 \, R_{\rm in}$). 
Using equation (5), we see that the best-fit value of $l$ 
is consistent with $M \sim 10^6 M_{\odot}$, as expected. 

We can obtain a stronger constraint on the central mass 
by inserting our best-fit value of $a$  
into equation (4): 

\begin{equation}
M \simeq 1.25 \times 10^6 \, l_{\rm Edd}^{-1/2} 
\left (\frac{\eta}{0.2} \right)^{1/2} \, M_{\odot}.
\end{equation}
For the high accretion rates observed in this system 
($0.3 \simlt l_{\rm Edd} \simlt 1$, see e.g. Puchnarewicz et al.\ 2001), 
and for $0.05 \simlt \eta \simlt 0.4$, we can constrain 
the mass of the central object to be
$0.6 \times 10^6 \simlt M \simlt 3 \times 10^6 M_{\odot}$.

A simple physical interpretation of $R_{\rm in}$ is that 
it corresponds to the radius of the innermost stable circular orbit 
around the central BH. However, 
the standard disk model is probably not applicable near 
the inner disk radius, and in any case it 
has to be modified with relativistic terms 
(e.g., Novikov \& Thorne 1973; Riffert \& Herold 1995). 
Therefore, our spectral 
fit near the peak of the blackbody component 
(at $\approx 0.1$--$0.5$ keV) can only be a general approximation. 
A more exact modelling is left to further work, and will 
require more precise observational data in the soft X-ray band. 
Nevertheless, we can check that 
our best-fit value of $R_{\rm in} = 0.56 \times 10^{12}$ cm 
is consistent with the range of values for $M$ derived before.  
If $M \leq 0.65 \times 10^6 M_{\odot}$, $R_{\rm in}$ 
is consistent with the innermost stable circular orbit 
of a Schwarzschild BH, while if 
$M > 1.5 \times 10^6 M_{\odot}$ we require 
a Kerr BH with angular parameter $a > 0.85$.

The outer disk radius is also constrained by our fit: 
we require $R_{\rm out} \simgt 5 \times 10^{16}$ cm 
$\approx 3 \times 10^5$ $GM/c^2$. If we keep all other parameters 
in Table 2 fixed, but assume a disk truncated at 
$R_{\rm out} = 10^{16}$ cm, we cannot reproduce 
the flat optical spectrum (curve marked ``a'' in Figure 2).

\begin{figure}
\begin{center}   
\epsfig{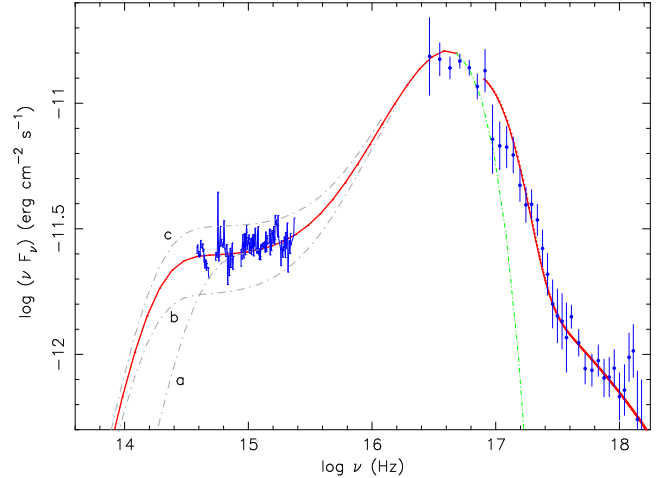}
\end{center}    
\caption{Broadband spectral fit to the WHT, HST, and {\it Beppo-SAX} 
observed fluxes ($E \leq 7$ keV). 
(The line-of-sight Galactic absorption has been removed.)
The thick solid line for $\log \nu < 16.7$ represents 
the low-energy part of an irradiated disk blackbody spectrum, continued as 
a dash-dotted line for $\log \nu > 16.7$. The thick solid line for 
$\log \nu > 16.9$ represents the X-ray spectral fit, physically 
interpreted as a disk blackbody modified by Comptonisation. 
The temperature at the inner edge of the disk 
$kT_{\rm in} = 0.1$ keV.  
The meaning of the curves labelled ``a'', ``b'' and ``c'' 
is discussed in the text.}
\end{figure}

\section{Further work: disk winds and line emission}

A large accretion disk heated by soft X-rays intercepted 
from the central source is likely to have 
a temperature-inversion layer at its surface 
(Wu et al.\ 2001). Therefore, we expect strong emission-line formation 
at large radii. In particular, in the case of \1034, the largest 
contribution to low-ionisation lines 
such as H$\alpha$ and H$\beta$ would come from radii $\simgt 10^{16}$ cm, 
where the Keplerian rotational velocities are 
$(GM/R)^{1/2} \simlt 1000$ km s$^{-1}$. If the Balmer lines 
are indeed produced near the disk surface, we expect 
full widths at half maximum $\sim 2(\sin i) (GM/R)^{1/2} 
\simlt 2000$ km s$^{-1}$. 
This is in agreement with the observed values of 1500 and 1800 km s$^{-1}$ 
for H$\beta$ and H$\alpha$ respectively (Puchnarewicz et al.\ 1995). 

Moreover, if the X-ray luminosity 
from the central source is $\simgt 0.1$ $L_{\rm Edd}$, we may expect 
the formation of a radiatively-driven accretion disk wind. 
Accurate modelling of the emission-line profiles 
may help us ascertain if higher-ionisation lines 
are formed in the wind, and determine the ionisation parameter 
of the emission regions. The narrow, single-peaked 
UV lines seen by Puchnarewicz et al.\ (1998) appear qualitatively consistent 
with the disk-wind model of Murray \& Chiang (1997). 
The presence of broad and narrow components in the UV lines 
may be explained with a broad component emitted near the irradiated 
disk surface at smaller radii, and a narrow component emitted in 
the photoionised wind. A full investigation of this issue 
is beyond the scope of this Letter and is left to further work.

\section{Conclusions}

We have fitted the optical to X-ray continuum spectrum of the 
narrow-line Seyfert 1 \1034 with an irradiated disk blackbody 
component (temperature 
$kT_{\rm in} \approx 0.1$ keV at the inner disk radius), 
and a power-law-like Comptonised component at higher energies. 
The disk component is much stronger and found at higher energies 
than in the average AGN; the spectral index 
of the power-law component is $\alpha \approx 1.6$ (steeper 
than the average value for an AGN); the observed luminosity is close to 
the Eddington limit. We infer a central mass of 
$0.6 \times 10^6 \simlt M \simlt 3 \times 10^6 M_{\odot}$.

The observed flat optical/UV spectrum is 
explained by our model as emission from the 
radiatively-heated outer part of the disk, at $R \simgt 80 \, R_{\rm in}$, 
where $T \sim R^{-1/2}$. Strong emission lines are likely to 
be formed near the irradiated surface of the outer disk, 
and in a radiatively-driven wind. Our model 
requires the disk to extend to $R_{\rm out} \approx 5 \times 10^{16}$ cm.
The observed flux and inner-disk temperature do not rule out 
the hypothesis of a face-on system, which would 
be in agreement with the low line-of-sight absorption 
seen in the X-ray spectra.

\end{document}